\documentclass[twocolumn,prl,superscriptaddress,letter]{revtex4}

\usepackage{graphicx}
\usepackage{amssymb}
\usepackage{amsfonts}

\DeclareGraphicsExtensions{.eps}

\begin{document}

\title{Quantum degeneracy of microcavity polaritons}

\author{A. Baas, J.-Ph. Karr,  M. Romanelli, A. Bramati, and E. Giacobino}
\affiliation{Laboratoire Kastler Brossel, Universit\'{e} Paris 6, \'{E}cole
Normale Sup\'{e}rieure et CNRS,\\
UPMC Case 74, 4 place Jussieu, 75252 Paris Cedex 05, France}

\begin{abstract}
We investigate experimentally one of the main features of a quantum fluid constituted by exciton polaritons in a
semiconductor microcavity , that is quantum degeneracy of a macroscopic fraction of the particles. We show that resonant
pumping allows to create a macroscopic population of polaritons in one quantum state. Furthermore we demonstrate that
parametric polariton scattering results in the transfer of a macroscopic population of polariton from one single quantum
state into another one. Finally we briefly outline a simple method which provides direct evidence of the first-order
spatial coherence of the transferred population.
\end{abstract}

\maketitle

In the past few years, the nature of quantum state of carriers in semiconductors and more specifically the possibility of
condensation has raised a growing interest \cite{Snoke}. Exciton polaritons in semiconductor microcavities appear as
promising candidates because they have a very small effective mass around the minimum energy state at $\mathbf{k=0}$
(where $\mathbf{k}$ is the wave vector in the plane of the layers), providing a large critical temperature for
condensation \cite{SemiSciTech,Rubo}. Bosonic stimulation of the relaxation towards the bottom of the dispersion curve
under nonresonant pumping has been reported by several groups \cite{Dang,Richard}. In order to demonstrate condensation,
it is crucial to give evidence of quantum degeneracy of polaritons. Here, we investigate this property in a very different
situation, where the lower polariton branch is pumped by a \emph{resonant} laser beam. The system is then far from
thermodynamical equilibrium; the polariton oscillation frequency is not fixed by an equation of state relating the
chemical potential to the polariton density and no condensation is expected to take place. The exciton polariton system is
however a quantum fluid which exhibits collective excitations \cite{Carusotto}.

We show in this paper that a macroscopic fraction of the particles in the system are in one single quantum state. To our
knowledge, there has been to date no complete experimental evidence of a polariton system being in a single quantum state.
Such evidence is necessarily related to coherence properties of the polariton field, which are transferred to the emitted
light field, due to its part-matter, part-light nature \cite{Ciuti,Karr04b}; this feature allows experimental access to
the polariton coherence through measurements on the emitted light field. Most experimental studies have focused on the
linewidth, angular divergence and phase coherence of the nonlinear emission \cite{Baumberg}. Second-order coherence in the
time domain has been studied theoretically \cite{Tassone,Schwendimann,Karr04b} and experimentally through the measurement
of the intensity autocorrelation function $g^{(2)}(\tau)$ of the emitted light \cite{Deng02}.

The reduction of the linewidth and angular divergence above threshold may indicate a reduction of the number of populated
polariton modes. However, such properties, related to the first-order coherence, are not in themselves sufficient proofs
of the single-mode nature of the polariton field. In the same way, a measurement of first-order temporal coherence does
not allow to discriminate between the light from a single-mode laser and the light from a spectrally filtered thermal
source \cite{Mandel}. In order to obtain a full characterization the measurement of the spatial second-order coherence
properties is a crucial one. The aim of this paper is to give an unambiguous experimental signature of polariton quantum
degeneracy in the regime of resonant pumping of the lower polariton branch, by studying the second-order spatial
coherence. We use specific methods from quantum optics to characterize the modes of the emitted light, which are directly
related to the properties of the modes of the polariton quantum fluid.

It has been shown recently that the measurement of intensity correlations in the transverse plane provides a criterion
allowing to characterize unambiguously single-mode and multimode quantum states of light \cite{Martinelli}. In this paper,
we apply this criterion to the light emitted by a polariton population around $\mathbf{k=0}$, created by polariton pair
scattering under resonant pumping of the lower polariton branch by a single-mode cw laser. Our results show that the
emitted light is in a single-mode quantum state, giving a proof of the quantum degeneracy of polaritons. We also
demonstrate first-order spatial coherence through the observation of interference fringes in the emitted field.

\begin{figure}[t]
\centerline{\includegraphics[clip=,width=8.5cm]{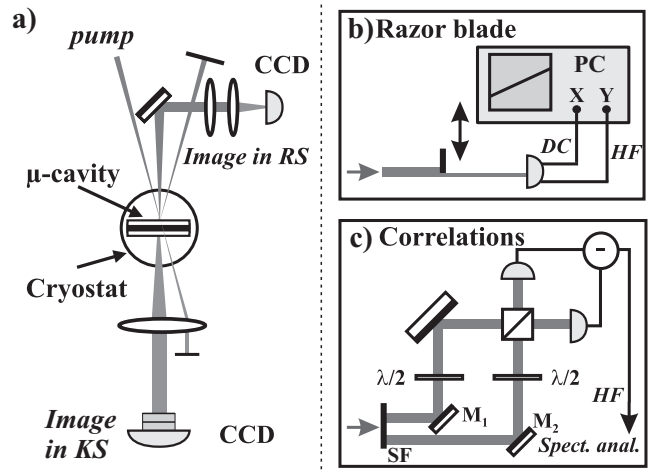}} \caption{a) Experimental setup. CCD camera for images in RS
(KS) of the emitted light in reflection (transmission). b) "Razor blade" setup. HF (DC): high (low) frequency part of the
signal. c) Intensity correlations set-up. Each of the mirrors $M_{1}$ and $M_{2}$ reflects the emission from one of the
regions selected by the spatial filter (SF) in KS (see also Fig.~\ref{Sorder}b).} \label{ExpSetUp}
\end{figure}

The microcavity sample is a high quality factor $2 \lambda$ cavity described in \cite{Houdre} with polariton linewidths in
the 100-200 $\mu$eV range. The pump laser is a single-mode tunable cw Ti:Saphir with a linewidth of 1 MHz, spatially
filtered through an optical fiber \cite{Baas04a} and focused on a 100$\mu$m spot (full width at half maximum of the
gaussian distribution) with a divergence of 0.6$^{\circ}$. All experiments are carried out with one $\sigma^{+}$ polarized
resonant pump beam and no probe beam. The left part of Fig.~\ref{ExpSetUp} shows the geometry of the experiment. Images of
the nonlinear emission around $\mathbf{k=0}$ are taken by CCD cameras in real-space (RS) (near field) and in
$\mathbf{k}$-space (KS) (far-field).

\begin{figure}[t]
\centerline{\includegraphics[clip=,width=8.5cm]{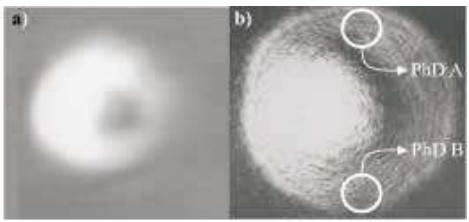}} \caption{Image of the non linear emission in RS (a) and in
$\mathbf{k}$-space (b) in the degenerate geometry \cite{Baas04a}. Cavity exciton detuning 0.3 meV and laser intensity 2.2
mW. In (a) the dark region in the laser spot corresponds to the polariton resonance. In (b) the polariton emission is seen
around the (brighter) reflected laser in $\mathbf{k}$-space, showing a larger angular divergence. The principle of the
intensity correlation measurement is schematically indicated.} \label{Sorder}
\end{figure}

We first consider the simplest case where the pump beam is at normal incidence and directly creates polaritons around
$\mathbf{k=0}$. The interest of investigating this configuration is to check the widely used assumption that a resonant
pump field creates polaritons in a single quantum state, provided the spot size and the pump divergence obey certain
conditions. The recent observation of complex transverse patterns in the emitted light in the nonlinear regime
\cite{Baas04a,Karr04a} raises some doubts about the single-mode character of the polariton population. Transverse
intensity correlation measurements allow to answer this question unambiguously. From a practical point of view, it is not
easy to analyze the properties of the polariton emission since it is in the same direction as the much more intense
reflected laser field. However, at high enough excitation density the polaritons are confined by nonlinear effects in a
zone that is smaller than the excitation spot \cite{Baas04a} (see the RS image of the excitation spot in Fig.~\ref{Sorder}
a). Because of diffraction, the emitted field has a larger angular divergence than the reflected laser (see the KS image
in Fig.~\ref{Sorder} b). The reflected laser can then be filtered out and we perform the intensity correlation
measurements on the remaining part of the emitted field.

Two methods will be used for the characterization of single-mode and multimode beams \cite{Martinelli}:

(i) For a beam in a single-mode quantum state, a partial intensity measurement (i.e. only a part of the beam is sent to
the photodetector) has the same effect as losses. This corresponds to the idea that in a single-mode quantum state,
photons are randomly distributed in the transverse plane. Such a behavior can be tested by "cutting" the beam
transversally with a razor blade and measuring the intensity noise as a function of intensity \cite{Poizat}. The noise
normalized to the shot noise limit should vary linearly with the intensity -the slope is respectively positive, negative
or equal to zero if the noise of the total beam is above, under or at the shot noise level. Any deviation from the linear
behavior would prove the multimode character of the beam.

(ii) If two separate parts of the single-mode beam of equal intensities are sent to two identical photodetectors $A$ and
$B$, the noise of the difference of the photocurrents $N_{A}-N_{B}$ is the shot noise, whatever the quantum state of the
beam -even for a beam with excess noise, when the noise measured by each detector is above the shot noise. There are no
correlations at the quantum level between $N_{A}$ and $N_{B}$, which again corresponds to the idea that photons are
randomly distributed in the transverse plane. Any deviation from the shot noise level would prove the multimode character
of the beam.

These methods only apply to the \textit{transverse} single-mode or multimode nature of the beam. It is assumed that only
one longitudinal cavity mode is present in the emission. For example, one can consider the case of a laser emitting on two
consecutive longitudinal cavity modes, but with the same transverse distribution described by a $TEM_{00}$ mode; it is
clear that the above-described experiments do not allow to evidence the multimode nature of such a beam. However in
microcavities the emission is obviously longitudinal single-mode, because other longitudinal modes are very far in energy.
In the same way it is assumed that only one polarization state is present in the emission. In our case, this condition can
be reached by using a $\sigma^{+}$-polarized pump beam, since it was shown that the signal emission is then also
$\sigma^{+}$ polarized \cite{Lagoudakis}.

In these experiments, a multimode beam can be identified unambiguously by a deviation from the expected behavior for a
single-mode beam. However, no necessary and sufficient condition can be given for the identification of a single-mode
quantum state, because the single-mode behavior has to be verified in \textit{all} transverse bases. But a convincing
indication can be given if the single-mode behavior is verified in a basis of transverse modes in which the field can be
expected to be multimode \cite{Martinelli}. Since the polariton modes are well approximated by planes waves defined by an
in-plane wave vector $\mathbf{k}$, the plane waves are a relevant basis to demonstrate the eventual multimode character of
the emission; in the following, all measurements are done in the far field.

\begin{figure}[t]
\centerline{\includegraphics[clip=,width=8.5cm]{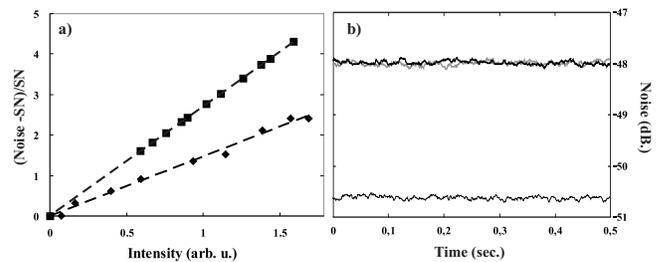}} \caption{a) Squares(diamonds): variations of the noise -HF
signal- of the nonlinear emission around $\mathbf{k}=0$ as a function of the detected intensity -DC signal-, in the
degenerate (non-degenerate) geometry. The shot noise (SN) is subtracted from the total noise, and the result is normalized
to the SN. Cavity-exciton detuning: respectively 0.3 and 1.2 meV. In the degenerate case, points at very low intensity are
missing because the noise was too small compared to dark noise. b) Noise power as observed on the screen of the spectrum
analyzer. Dark line: noise of the intensity difference, perfectly superimposed with the shot noise level (gray line). Dark
noise (clear gray line) is $2.6$ dB below the SN level.} \label{Sorderbis}
\end{figure}

The curve in Fig.~\ref{Sorderbis} a) shows the results of the "razor blade" experiment. The linear variation of the
normalized noise as a function of intensity indicates the single-mode character of the emitted light ; the slope is
positive because the beam has some excess noise.

This result was confirmed by the measurement of intensity correlations in the transverse plane. The experimental setup is
shown in Fig.\ref{ExpSetUp} c). Two parts of the emitted beam are selected (as shown in Fig.~\ref{Sorder} b) by placing a
screen with two small holes in $\mathbf{k}$-space. The light transmitted by each of the holes is sent to a detector by
removing the $\lambda/2$ waveplates and the high-frequency part of the photocurrent difference is sent to a spectrum
analyzer. The polarizing cube, together with the $\lambda/2$ waveplates, allows to measure the shot noise reference by
separating each beam into two parts of equal intensities incident on each detector and making the difference of the
photocurrents \cite{Bachor}. An example of our results is shown in Fig.~\ref{Sorderbis} b). The noise of the intensity
difference is the shot noise level, whereas each beam has an excess noise of more than $30 \%$. This experiment was
repeated for different positions of the holes in $\mathbf{k}$-space, with identical results.

These results are a very strong indication that the emission around $\mathbf{k=0}$ originates from one and the same
polariton quantum state, i.e. the k=0 state populated by resonant pumping. In particular, it implies that non-degenerate
processes such as the pair scattering $\{ \mathbf{0,0} \} \rightarrow \{ \mathbf{k,-k} \}$ play a negligible role and
validates the single-mode treatment adopted in \cite{Schwendimann,Baas04a,Karr04a, Ciuti}. The larger divergence of the
nonlinear emission with respect to that of the pump beam is due to diffraction effects and is not associated with a
polariton population in other modes.

We now consider a pump at the so-called "magic angle" with a power slightly above the parametric oscillation threshold,
where a bright signal beam is emitted around $\mathbf{k=0}$ \cite{Savvidis}. Also in this configuration, complex
transverse patterns have been observed in the emitted light around $\mathbf{k=0}$ \cite{Houdre,Baas04b}, raising questions
about its single-mode character. Again, intensity correlation measurements provide an unambiguous answer.

Fig.~\ref{Sorderbis} a) shows the result of a "razor blade"-type experiment (Fig.~\ref{ExpSetUp} b) on the light emitted
around $\mathbf{k=0}$ in reflection, the reflected pump beam being carefully filtered out. The emitted beam is cut by a
razor blade in $\mathbf{k}$-space. The variation of the normalized noise as a function of the intensity is linear within
experimental accuracy, the positive slope meaning that the total beam has excess noise. This experiment demonstrates that
only one signal-idler pair oscillates, as can be expected from the analogy with optical parametric oscillators
\cite{fabrecras}. It validates the widely used model involving only three polariton $\mathbf{k}$ states \cite{Ciuti},
which successfully accounts for most experimental observations \cite{SemiSciTech}. In other words, polariton pair
scattering makes it possible to create a macroscopic population of polariton in one single quantum state distinct from the
pumped mode. It would be very interesting to make the same kind of measurement below threshold - which requires more
sensitive photodetectors than the one used here-, since many signal-idler pairs are then involved and evidence for a
multimode quantum state should be observed \cite{Langbein}.

The single-mode nature of the signal polariton field implies phase coherence, i.e. first-order spatial coherence over the
whole emission zone. In order to check this property, we use an original set-up directly revealing the phase coherence of
the signal field. The idea is to superimpose the light fields emitted by two separate spots of the sample, sufficiently
small to have a significant divergence so that the emitted fields overlap in $\mathbf{k}$-space. This is analogous to
Young's double slit experiment, where the slits are put directly on the surface of the light source.

\begin{figure}[t]
\centerline{\includegraphics[clip=,width=8.5cm]{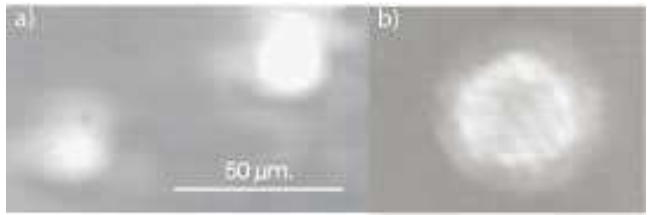}} \caption{Image of the light emitted in a cone of $2^{\circ}$
around the normal direction a)in RS in reflection ; b) in $\mathbf{k}$-space in transmission. Cavity-exciton detuning: 1.3
meV. Laser intensity: 2.8 mW (around 1.5 times the threshold intensity).} \label{Forder}
\end{figure}

Fig.~\ref{Forder} a) shows a RS image of the sample surface, after filtration in KS of the light emitted in a cone of
$2^{\circ}$ around the normal direction, which allows to get rid of the pump laser at the "magic angle" (about
$12^{\circ}$). It can be seen that the signal field is emitted by two separate spots of a few microns in diameter and
separated by $l \simeq$ 70 $\mu$m. The cavity-exciton detuning and excitation intensity have been chosen precisely to
reach this situation ,which is caused by the interplay between nonlinear and disorder effects, so as to make the
implementation of our Young's experiment easier. However, the separation of the polariton field in two spots is not in
principle necessary for the experiment; alternatively, it is possible to make an image of the surface of the sample and
place two slits in the image plane.

The corresponding image in KS presents interference fringes (Fig.~\ref{Forder} b). We are confident that they are not due
to a self-interference of the laser, because the laser light has been filtered out. We sent the emitted field through a
monochromator and found no component at the laser wavelength. Moreover, the fringes are perpendicular to the line joining
the two spots and the fringe separation is in full agreement with a simple two-slits type calculation: $i=\lambda f//l$,
with f the focal of the lens. This experiment reveals a significant degree of first-order spatial coherence of the signal
polaritons at a scale of 70 $\mu$m. The fact that interference is observed in cw regime confirms, in addition, that the
beams emitted by the two spots originates from the same polariton mode (in spite of the complex shape of the active region
in RS). Indeed, two distinct modes would interfere during a time limited by the coherence time $T_{2}$, in the same way as
the emissions from two independent lasers \cite{Mandel}.

This result is in full agreement with the single-mode properties discussed above. It also illustrates a direct method of
measurement of first-order spatial coherence and gives an order of magnitude for the coherence length of the polaritons
created by the parametric process, which can be compared to the coherence length of polaritons in the linear regime
\cite{Tartakovskii00}.

We have presented an experimental study of quantum degeneracy of microcavity polaritons, relying on the measurement of
intensity correlations in the transverse plane. We have shown that (i) resonant pumping allows to create a macroscopic
population of polaritons in a single quantum state; (ii) in the case of non-degenerate interaction, the polariton pair
scattering gives rise to a single-mode quantum state for the polariton population. These results are in agreement the
description of the system as a quantum fluid in which only a few collective excitation modes are populated
\cite{Carusotto}. The experimental methods that we have used are very promising for the characterization of quantum
degeneracy \cite{Butov,Richard}, also under nonresonant pumping.

We acknowledge enlightening discussions with I. Carusotto, C. Ciuti, O. El Da\"{\i}f,  C. Fabre, T. Guillet, N. Treps, and
J. Tignon. We are very grateful to R. Houdr\'{e} for providing us with the microcavity sample.


\begin{thebibliography}{99}

\bibitem{Snoke} S. A. Moskalenko and D. W. Snoke, \textit{Bose-Einstein Condensation of Excitons and Biexcitons and
Coherent Nonlinear Optics with Excitons} (Cambridge University Press, Cambridge, 2000).
\bibitem{SemiSciTech} Semicond. Sci. Technol. \textbf{18}, S279-S434, special issue on semiconductor microcavities,
edited by J. J. Baumberg and L. Vina (2003).
\bibitem{Rubo} Y. G. Rubo \textit{et al.}, Phys. Rev. Lett. \textbf{91}, 156403 (2003); J. Keeling \textit{et al.},
Phys. Rev. Lett. \textbf{93}, 226403 (2004).
\bibitem{Dang} D. Le Si Dang \textit{et al.}, Phys. Rev. Lett. \textbf{81}, 3920 (1998); P. Senellart \textit{et al.},
Phys. Rev. B \textbf{62}, R16263 (2000); R. Butt\'{e} \textit{et al.}, Phys. Rev. B \textbf{65}, 205310 (2002).
\bibitem{Richard}M. Richard \textit{et al.}, Phys. Rev. B \textbf{72}, 201301 (2005).
\bibitem{Carusotto} I. Carusotto and C. Ciuti,Phys. Rev. Lett. \textbf{93}, 166401 (2004);
\bibitem{Ciuti} C. Ciuti \textit{et al.}, Phys. Rev. B \textbf{62}, R4825 (2000).
\bibitem{Karr04b} J. Ph. Karr \textit{et al.},Phys. Rev. A \textbf{69}, 063807 (2004).
\bibitem{Baumberg} J. J. Baumberg \textit{et al.}, Phys. Rev. B \textbf{62}, R16247 (2000);
G. Messin \textit{et al.}, Phys. Rev. Lett. \textbf{87}, 127403 (2001); A. Huynh \textit{et al.}, Phys. Rev. Lett.
\textbf{90}, 106401 (2003); S. Kunderman \textit{et al.}, Phys. Rev. Lett \textbf{91}, 156403 (2003).
\bibitem{Tassone} F. Tassone and Y. Yamamoto, Phys. Rev. A \textbf{62}, 063809 (2000); F. P. Laussy \textit{et al.},
Phys. Rev. Lett. \textbf{93}, 016402 (2004); D. Sarchi and V. Savona, cond-mat/0411084.
\bibitem{Schwendimann} P. Schwendimann \textit{et al.}, Phys. Rev. B \textbf{68}, 165324 (2003).
\bibitem{Deng02} H. Deng \textit{et al.}, Science \textbf{298}, 199 (2002).
\bibitem{Mandel} L. Mandel and E. Wolf, \textit{Optical Coherence and Quantum Optics} (Cambridge University Press,
Cambridge, 1995).
\bibitem{Martinelli} M. Martinelli \textit{et al.}, Phys. Rev. A \textbf{67}, 023808 (2003).
\bibitem{Houdre} R. Houdr\'{e} \textit{et al.}, Phys. Rev. Lett. \textbf{85}, 2793 (2000).
\bibitem{Baas04a} A. Baas \textit{et al.}, Phys. Rev. A \textbf{69}, 023809 (2004).
\bibitem{Karr04a} J. Ph. Karr \textit{et al.}, Phys. Rev. A \textbf{69}, 031802(R) (2004).
\bibitem{Poizat} J.-Ph. Poizat \textit{et al.}, J. Opt. Soc. Am. B \textbf{15}, 1757 (1998); J.-P. Hermier
\textit{et al.}, J. Opt. Soc. Am. B \textbf{16}, 2140 (1999).
\bibitem{Lagoudakis} P. G. Lagoudakis \textit{et al.}, Phys. Rev. B \textbf{65}, 161310 (2002).
\bibitem{Bachor} H.A. Bachor, \textit{A Guide to Experiments in Quantum Optics}, Wiley, Weinheim, pp. 181-183 (1998).
\bibitem{Savvidis} P. G. Savvidis \textit{et al.}, Phys. Rev. Lett. \textbf{84}, 1547 (2000); R. M. Stevenson
\textit{et al.}, Phys. Rev. Lett. \textbf{85}, 3680 (2000); J. Erland \textit{et al.}, Phys. Rev. Lett.
\textbf{86}, 5791 (2001); M. Saba \textit{et al.} Nature, \textbf{414}, 731 (2001); A. I. Tartakovskii \textit{et al.},
Phys. Rev. B \textbf{65}, 081308 (2002).
\bibitem{Baas04b} A. Baas \textit{et al.}, Phys. Rev. B \textbf{70} 161307(R) (2004).
\bibitem{fabrecras} C. Fabre \textit{et al.}, C. R. Acad. Sci. Paris 1 IV, 553 (2000).
\bibitem{Langbein} W. Langbein \textit{et al.}, Phys. Rev. B \textbf{70}, 205301 (2004).
\bibitem{Tartakovskii00} A. I. Tartakovskii \textit{et al.}, Phys. Stat. Sol. B \textbf{221}, 163 (2000).
\bibitem{Butov} L. V. Butov \textit{et al.}, Nature \textbf{417}, 47 (2003); C. W. Lai \textit{et al.},
Science \textbf{303}, 503 (2004).

\end{thebibliography}
\end{document}